\documentclass[12pt]{article}
\usepackage{amsmath,amsfonts}
\def\sloppy{\tolerance=100000\hfuzz=\maxdimen\vfuzz=\maxdimen}
\def \be {\begin{equation}}
\def \ee  {\end{equation}}
\def \beqs {\begin{eqnarray}}
\def \eeqs {\end{eqnarray}}
\def\sqr#1#2{{\vcenter{\vbox{\hrule height.#2pt
\hbox{\vrule width.#2pt height#1pt \kern#1pt
\vrule width.#2pt}\hrule height.#2pt}}}}

\def\Xint#1{\mathchoice
   {\XXint\displaystyle\textstyle{#1}}%
   {\XXint\textstyle\scriptstyle{#1}}%
   {\XXint\scriptstyle\scriptscriptstyle{#1}}%
   {\XXint\scriptscriptstyle\scriptscriptstyle{#1}}%
   \!\int}
\def\XXint#1#2#3{{\setbox0=\hbox{$#1{#2#3}{\int}$}
     \vcenter{\hbox{$#2#3$}}\kern-.5\wd0}}

\def\dashint{\Xint-}

\def \tr {{\rm tr}}

\begin{document}
\fontfamily{cmr}\fontsize{11pt}{14pt}\selectfont
\def \CMP {{Commun. Math. Phys.}}
\def \PRL {{Phys. Rev. Lett.}}
\def \PL {{Phys. Lett.}}
\def \NPBProc {{Nucl. Phys. B (Proc. Suppl.)}}
\def \NP {{Nucl. Phys.}}
\def \RMP {{Rev. Mod. Phys.}}
\def \JGP {{J. Geom. Phys.}}
\def \CQG {{Class. Quant. Grav.}}
\def \MPL {{Mod. Phys. Lett.}}
\def \IJMP {{ Int. J. Mod. Phys.}}
\def \JHEP {{JHEP}}
\def \PR {{Phys. Rev.}}
\def \JMP {{J. Math. Phys.}}
\def \GRG{{Gen. Rel. Grav.}}
\begin{titlepage}
\null\vspace{-62pt} \pagestyle{empty}
\begin{center}
\rightline{CCNY-HEP-12/3}
\rightline{February 2012}
\vspace{1truein} {\Large
Two-dimensional Born-Infeld gauge theory: spectrum, \\
string picture and large-$N$ phase transition}\\
\vspace{6pt}
\vskip .1in
{\Large \bfseries  ~}\\
\vskip .1in
{\Large\bfseries ~}\\
{\large Alexios P. Polychronakos}\\
\vskip .2in
{\itshape Physics Department\\
City College of the CUNY\\
160 Convent Avenue, New York, NY 10031}\\
\vskip .1in
\begin{tabular}{r l}
E-mail:
&{\fontfamily{cmtt}\fontsize{11pt}{11pt}\selectfont alexios@sci.ccny.cuny.edu}\\
\end{tabular}

\fontfamily{cmr}\fontsize{11pt}{15pt}\selectfont
\vspace{.8in}
\centerline{\large\bf Abstract}
\end{center}
We analyze $U(N)$ Born-Infeld gauge theory in two spacetime dimensions. We derive the exact energy spectrum
on the circle and show that it reduces to $N$ relativistic fermions on a dual space. This contrasts to the Yang-Mills
case that reduces to nonrelativistic fermions. The theory admits a string theory interpretation, analogous to
the one for ordinary Yang-Mills, but with higher order string interactions. We also demonstrate that the partition function on
the sphere exhibits a large-$N$ phase transition in the area and calculate the critical area. The limit in which the dimensionless
coupling of the theory goes to zero corresponds to massless fermions, admits a perturbatively exact free string interpretation
and exhibits no phase transition.
\end{titlepage}
\section{Introduction}

Two-dimensional gauge theory is special in that local gauge field excitations are absent and only global
variables (holonomies) remain as physically relevant degrees of freedom. As such, it is usually amenable to an 
exact treatment and provides a convenient testing ground for conjectures on the properties of gauge theory.
In particular, large-$N$ properties, such as the equivalence of gauge theory and string theory \cite{Hoo,Wil,Sak,BBHP} and the
analyticity of the strong coupling expansion can be directly probed.

The case of ordinary Yang-Mills theory has been analyzed exhaustively. The standard plaquette action has been
shown to exhibit a large-$N$ phase transition, leading to a nonanalyticity of the strong coupling expansion
below a critical coupling \cite{GrWi,Wad}. The true fixed point of the plaquette action was shown to be the so-called heat kernel
action \cite{Mig,Rus}, which gives analytic results and allows for the exact calculation of Wilson loop expectation values and of the
partition function in terms of infinite sums over representations of the gauge group \cite{KaKo,Kaz}.

A particularly attractive feature of two-dimensional Yang Mills theory
is its exact description as a string theory. This was shown both at
the level of Wilson loop expectation values, which admit an
interpretation as string coverings of the loop area with
(taut) string worldsheets of various windings \cite{KaKo,Kaz, Kos}, and the partition
function, which can be calculated in terms of wrappings of the
spacetime manifold with string worldsheets of various windings,
including string splitting and joining interactions
\cite{Gro,Min,GrTa}. As usual,
$1/N$ plays the role of the string coupling constant. Nonperturbative
effects of order $exp(-N)$ also appear, with string winding number
being conserved only modulo $N$. These results hold without the
benefit of supersymmetry and are based on pure group theory of the
gauge manifold. Further, the partition function on spacetimes of
spherical topology exhibits a large-$N$ phase transition in the
spacetime area, going from a strongly coupled (stringy) phase for
large area to a weakly coupled phase for smaller area \cite{DoKa}.

Two-dimensional Yang-Mills theory also admits alternative and
equivalent descriptions as a collection of free nonrelativistic
fermions \cite{MiPo1,MiPo2}, a gauged unitary matrix model \cite{Pol1}
and a $c=1$ collective field theory \cite{DaJe}.
Although the interconnection between these descriptions is known,
gauge theory presents a concrete physical realization and allows for
a convenient formulation of related string or many-body problems.
For instance, two-dimensional Yang-Mills is related to the
Sutherland model \cite{GoNe}, and generalized interacting
Calogero-Sutherland type integrable models of particles with
internal degrees of freedom can be obtained and solved in terms
of Yang-Mills theory on the cylinder with one or more Wilson
loop insertions \cite{MiPo3}.

Born-Infeld electrodynamics was introduced originally as an attempt
to provide a regularization of the short-distance singularity problem
of standard electrodynamics while preserving relativistic invariance,
at the price of a nonpolynomial action. In more recent contexts,
Born-Infeld actions often arise for the gauge fields in string and
brane theory (see, e.g., \cite{FrTs,Lei}). 
The obvious question is, then, whether such theories
in two dimensions are amenable to solution, admit a string
interpretation and share the qualitative and quantitative
features of standard Yang-Mills theory.

In this paper we analyze two-dimensional Born-Infeld theory and
address the above questions. Unlike Yang-Mills theory, which has a
unique fixed point, Born-Infeld theory can flow to various
inequivalet theories according to the exact renormalization and
ordering of its nonpolynomial terms. Under the most ``natural"
choice, the theory becomes equivalent to a set of {\it relativistic}
fermions, providing a nice generalization of the corresponding
Yang-Mills result of nonrelativistic fermions. A string interpretation
is still possible but involves higher-order string interactions.
On the sphere, a large-$N$ third-order phase transition is also present
with the same qualitative features as in Yang-Mills.

The limit in which the dimensionless
coupling of the theory goes to zero is particularly interesting:
the corresponding fermion picture involves massless particles and
admits a perturbatively exact free string interpretation. Further,
the would-be phase transition on the sphere disappears, the theory
being always in a nominally strongly coupled phase, with a smooth
crossover from a true stringy phase to an effectively weakly
coupled phase.

\section{Classical analysis of the system}

Nonabelian Born-Infeld actions are, in general, not unique even
at the classical level due to ordering ambiguities \cite{Tse}. 
Specifically, the determinant over
spacetime indices in their lagrangian involves the product of the components of the field strength tensor, which are matrices
and their ordering matters. The unique exception is two-dimensioal theory, where the field strength has a unique nonzero
component and there in no ambiguity.

We shall consider $U(N)$ gauge theory on a cylindrical spacetime manifold of spatial period $L$. The field strength is
\be
F_{\mu \nu} = \partial_\mu A_\nu - \partial_\nu A_\mu + i [ A_\mu , A_\nu ]~,~~~ \mu \,, \, \nu = t,x
\ee
The lagrangian of the theory can be written as
\be
{\cal L} = -\lambda \, \tr \sqrt{ - \det ( \eta_{\mu \nu} + F_{\mu \nu} /T)}
\ee
where the trace is over $U(N)$ matrices while the determinant is over spactime indices. $\lambda$ and $T$ are dimensionful constants,
playing the role of brane tension and string tension, respectively.
Classically, $\lambda$
is irrelevant and can be set equal to $T$. Quantum mechanically,
however, it is a relevant parameter, and the ratio $\lambda / T$
becomes a dimensionless coupling constant for the theory.
In particular, the ``tensionless"
limit $\lambda /T \to 0$ is particularly interesting as will be
shown in the sequel.

From now of we shall denote by
\be
F= F_{tx} = \partial_t A_x - \partial_x A_t + i [ A_t , A_x ]
\ee
the unique nonzero component of the field strength, in terms of which the action becomes
\be
S = \int d^2 x \, {\cal L} = -\lambda \int d^2x ~
\tr \sqrt{1 - F^2 / T^2}
\ee
In the limit
$|F| \ll T$ the above action becomes (up to an irrelevant 
additive constant)
\be
S \simeq \int d^2x \, \frac{\lambda}{2T^2} \, \tr F^2 
= \int d^2x \, \frac{\lambda}{4T^2} \, \tr F_{\mu \nu}^2
\equiv \int d^2x \, \frac{1}{4g^2} \, \tr F_{\mu \nu}^2
\ee
which identifies the Yang-Mills coupling in that limit as
\be
g^2 = \frac{T^2}{\lambda}
\label{YMg}
\ee
At the large $N$ limit the 't Hooft scaling of the YM coupling is
\be
g^2 = \frac{g_o^2}{N}
\label{YMN}
\ee

It is usful to recast the theory in a first-order formalism. To this end, we define the gauge-covariant momentum
\be
B = \frac{\delta {\cal L}}{\delta F} = \frac{\lambda F}{T^2 \sqrt{ 1 - F^2 /T^2}}
\label{B}
\ee
in terms of which the hamiltonian density is
\be
{\cal H} = \tr (B F - {\cal L}) = \tr \sqrt{\lambda ^2 + T^2 B^2}
\ee
while the action becomes
\be
S = \int d^2x \, \tr \left( BF - \sqrt{\lambda ^2 + T^2 B^2} \right)
\ee
The variation in $B$ of this action yields its defining equation (\ref{B}), while the variation of $A_\mu$ gives the
Gauss law contraint and equation of motion, respectively
\be
D_x B = 0 ~,~~~ D_t B = 0
\ee
We can use gauge invariance to put $A_t = 0$, provided we impose the Gauss law as a constraint.
In terms of the unique gauge field $A_x \equiv A$ the action is
\be
S = \int dt \int_0^L dx \, \tr \left( B {\dot A} - \sqrt{\lambda ^2 + T^2 B^2} \right)
\label{act}
\ee
while the Gauss law constraint remains
\be
\partial B + i [A , B ] = 0
\ee
where overdot and $\partial$ stand for time and space differentiation respectively.

The above theory has no local excitations and its only degree of freedom is the nontrivial holonolmy (Wilson loop)
around the spatial circle. To reduce the theory to its degrees of freedom, we proceed in close analogy to \cite{MiPo1}.
We define the spatial open Wilson loop
\be
W_{a,b} = P e^{i \int_a^b A dx} ~,~~~ W_{b,a} = W_{a,b}^{-1}
\ee
and consider the dressed momentum
\be
\Pi (x) = W_{0,x} B(x) W_{x,L}
\ee
Upon differentiating $\Pi$ with respect to $x$ and using the Gauss law we get
\be
\partial \Pi = W_{0,x} \Big( \partial B(x) + i [A(x) , B(x) ] \Big) W_{x,L} = 0
\ee
So $\Pi$ is spatially constant and from $\Pi (x) = \Pi (0)$ we obtain
\be
B(x) = W_{x,0} B(0) W_{0,x} = W_{0,x}^{-1} B(0) W_{0,x}
\ee
The kinetic term in the action (\ref{act}) can be expressed as
\be
K = \int dt dx \, \tr(B {\dot A} ) = \int dt \, \tr\left[ B(0) \int dx W_{0,x} {\dot A} W_{x,0} \right]
\ee
The time derivative of the full Wilson loop $W_{0,L}$, on the other hand, is
\be
{\dot W}_{0,L} = \int dx \, W_{0,x} i {\dot A}(x) W_{x,L} = 
i \int dx \, W_{0,x} {\dot A}(x) W_{x,0} W_{0,L}
\ee
and comparing with the kinetic term above we find
\be
K = -i \int dt \, \tr \Big( B(0) {\dot W}_{0,L} W_{0,L}^{-1} \Big)
\ee
Similarly, the hamiltonian becomes
\be
H = \int dx \, \tr  \sqrt{\lambda ^2 + T^2 B(x)^2 } = \int dx \, 
\tr \sqrt{\lambda ^2 + T^2 W_{0,x}^{-1} B(0)^2 W_{0,x}}
= L \tr \, \sqrt{\lambda ^2 + T^2 B(0)^2 }
\ee
due to the cyclicity of trace. So the full action can be expressed in terms of the space-independent fields
$P \equiv B(0)$ and $U \equiv W_{0,L}$ as
\be
S = \int dt \, \tr \left( -i P {\dot U} U^{-1} -  L \sqrt{\lambda ^2 + T^2 P^2} \right)
\ee
In addition, there is one residual Gauss law constraint: the periodicity condition $B(0) = B(L)$ gives
\be
P = U P U^{-1} ~~~{\rm or}~~~ [P,U] = 0
\label{GG}
\ee
In conclusion, we see that the theory reduces to a matrix model for the unitary matrix $U$ and the hermitian
matrix $P$ that plays the role of its canonical right-momentum. This model differs from the traditional unitary
matrix model in that its kinetic energy is a nontrivial (non-quadratic) function of the canonical momentum.

The above matrix model can be further reduced to noninteracting particles upon use of the Gauss constraint (\ref{GG}).
Classically, upon use of the equations of motion, $P$ is a function of ${\dot U} U^{-1}$, so the constraint implies
\be
[ P , U ] = [ U , {\dot U} ] =0
\ee
This means that $P$ and $U$ can be simultaneously diagonalized with a time-independent unitary transformation, 
reducing them to their eigenvalues $p_n$ and $e^{i \theta_n}$ respectively. To make this more explicit, we
write $P$ and $U$ in the basis where $U$ is diagonal as
\be
U = V \Lambda V^{-1} ~,~~~ P = V ( p + Q ) V^{-1}
\ee
with $V$ the diagonalizer of $U$, $\Lambda$ and $p$ diagonal and $Q$ off-diagonal, that is
\be
\Lambda = diag \{ e^{i \theta_n} \} ~,~~~ p = diag \{ p_n \} ~,~~ Q_{nn} = 0 ~\rm{(no~sum~in}~n\rm{)}
\ee
The canonical (time derivative) term in the action in terms of the above variables is
\be
-i \, \tr \left( P {\dot U} U^{-1} \right) =  
\sum_{n=1}^N  p_n {\dot \theta}_n + \sum_{n,m} 
\left( 1 - e^{i(\theta_n - \theta_m )} \right) Q_{nm} \left( {\dot V} V^{-1} \right)_{mn}
\ee
We see that the eigenvalues $\theta_n$ are canonically conjugate to the diagonal elements $p_n$
of $P$, while $Q - \Lambda Q \Lambda^{-1}$ is the right-momentum of the angular part of $U$ (the
dagonalizer $V$). The Gauss law (\ref{GG}) implies $Q=0$. This is a gauge constraint and must therefore
be complemented by gauge fixing the coordinates corresponding to $Q$, that is $V$. Setting $V=1$
we are left with a set on $N$ coordinates on the unit circle and the their canonical momenta.
The reduced action becomes
\be
S = \int dt \, \sum_{n=1}^N \left( p_x {\dot \theta}_n -  L \sqrt{\lambda ^2 + T^2 p_n^2} \right)
\ee
and describes a set of nonintercting particles with a relativistic energy-momentum relation:
\be
E = L \sqrt{\lambda ^2 + T^2 p^2}
\ee
with $LT$ and $\lambda/LT^2$ playing the role of the speed of light and the particle's mass respectively.

\section{Quantization}

Quantum mechanically the story is similar, with some additional twists. 
The wavefunction $\Phi(U)$ is a function of the matrix elements of $U$. 
From the canonical structute of the action, 
$P$ generates right-multiplications of $U$ by unitary matrices, while $-U P U^{-1}$ generates left-multiplications.
Their sum $P - U P U^{-1}$ generates the conjugation
\be
U \to V^{-1} U V
\ee
The Gauss constraint (\ref{GG}) implies that wavefunctions are invariant under unitary conjugations of $U$
and therefore depend only on the eignevalues of $U$. As usual, the change of variables
from $U$ to its eigenvalues $e^{i \theta_n}$ and the angular variables $V$ involves the Jacobian of the
transformation $J = |\Delta|^2$, where $\Delta$ is the modified Vandermonde factor
\be
\Delta (\theta ) = \prod_{n<m} \sin \frac{\theta_n - \theta_m}{2}
\label{VM}
\ee
The Jacobian can be absorbed by incorporating one factor of $\Delta (\theta )$ in the wavefunction, rendering
the measure in $\theta_n$ flat. The original wavefunction $\Phi (\theta )$ was symmetric under permutation
of $\theta_n$, so the new wavefunction
\be
\Psi (\theta ) =  \prod_{n<m} \sin \frac{\theta_n - \theta_m}{2} 
\, \Phi (\theta )
\ee
becomes fermionic. (This is the famous fermionization of the eigenvalues of a matrix model, whch holds
irrespective of its action.)

The spectrum is evaluated by diagonalizing the hamiltonian
\be
H = L \tr\sqrt{\lambda ^2 + T^2 P^2} = L \lambda \sum_{n=0}^\infty c_n \left( \frac{T}{\lambda} \right)^{2n}
\tr P^{2n}
\label{expand}
\ee
with $c_n$ the Taylor expansion coefficients of the square root. Since $P$ generates unitary transformations
(left-multiplications of $U$) it satisfies the $U(N)$ algerba. The quantum commutation relations of its matrix elements
read
\be
[ P_{mn} , P_{kl} ] = i \left( P_{ml} \delta_{kn} - P_{kn} \delta_{ml} \right)
\ee
and $\tr P^n$ is the $n$-th Casimir operator for $U(N)$ ($\tr P$ being the $U(1)$ charge). These are diagonalized
on states that are irreducible representations (irreps) of the above algebra. Given that states must also be singlets under
conjugation of $U$, they are the characters of the representations. The fermionic eigenstates are
\be
\Psi_{_R} = \Delta (\theta ) \chi_{_R} (U) = \Delta (\theta ) \tr_{_R} U_{_R}
\ee
with $R$ an irrep of $U(N)$ and $U_{_R}$ the $R$-matrix representation of the element $U$. For $U(N)$,
the Casimirs of order larger than $N$ are not independent but they are still diagonal on irreps.

The above also leads to the result that the energy
states of the theory are simply free states of $N$ fermions on the unit circle determined by their momenta.
Specifically, they are given by the Slater determinant
\be
\Psi (\theta ) = \det_{kn} ( e^{i p_k \theta_n} )
\label{SL}
\ee
The fermion momenta $p_k$ can be ordered as $p_{k+1} <  p_k$. Since $\theta_n$ have period
$2\pi$, the momenta are quantized to integer steps plus, perhaps, a constant shift. The shift is determined
by the properties of Vandermonde factor (\ref{VM}), which is periodic
for odd $N$ and antiperiodic for even $N$.
So the momenta $p_k$ are quantized to (half) integers for (even) odd $N$. The ground state is in both cases
\be
p_{k,o} = \frac{N+1}{2} - k =  \left\{ \frac{N-1}{2} , \, \frac{N-1}{2} - 1 , \, \dots \, , -\frac{N-1}{2} \right\}
\label{pgr}
\ee
representing a Fermi sea symmetric around $p=0$ with Fermi momentum $p_{_F} = (N-1)/2$.
The Slater determinant (\ref{SL}) for the ground state is exactly the Vandermonde factor, 
$\Psi_o = \Delta$, leading to the bosonic ground state 
$\Phi_o (U) = 1$, that is, the singlet.

The relation of the momenta $p_k$ with the irreps they correspond is standard: the excitation
of each momentum from its ground state
\be
l_k = p_k - p_{k,o} = p_k + k - \frac{N+1}{2}
\ee
satisfying $l_{k+1} \le l_k$, represents the length of the $k$-th row in the Young tableau of the irrep.
The total number of boxes $\sum l = \sum p$ is the $U(1)$ charge. Negative lengths correspond to
conjugate irreps and can be turned positive by adding a number of columns of length $N$, that is,
by increasing the $U(1)$ charge by multiples of $N$. Note that for our $U(N)$ matrix model the
$U(1)$ charge $Q$ and the $SU(N)$ irrep are correlated in that the $Z_N$ charge $Z$ is common to both
and thus $Z = exp(i Q)$.

Since the problem reduces to free particles classically, we expect $P$ to act essentially as the diagonal momenta
$p_n$ conjugate to $\theta_n$, that is $-i\partial / \partial \theta_n$. This is, indeed, true for the first two
Casimirs. $\tr P$ reduces to the total momentum of the particles
\be
\tr P = \sum_n p_n = -i \sum_n \frac{\partial}{\partial \theta_n}
\ee
since the $U(1)$ charge is just an overall shift of the eigenvalues of $U$. $\tr P^2$ is essentially their
quadratic kinetic energy:
\be
\tr P^2 = \sum_n p_n^2 = - \sum_n \frac{\partial^2}{\partial \theta_n^2} - C
\ee
with $C$ a c-number subtracting the ground-state value of the right hand side operator.
In fact, $\tr P^2$ is the Laplacian on the group $U(N)$ and it is known to reduce to the above expression
when acting on Schur (conjugation-invariant) states.

The situation with higher Casimirs is subtler. In fact, $\tr P^n$ does {\it not} reduce to $\sum p_k^n$
for $n>2$, but involves also polynomials in lower-power sums of $p_k$
(see, e.g., \cite{Dou} and references therein). 
In the classical limit, that is,
for $|p_k | \gg N$, the two expressions must agree. So we have
\be
\tr P^n = \sum_k p_k^2 + \rm{lower~order~terms}
\label{Ord}
\ee
This can be viewed as a quantum effect arising from ordering issues in the field theory. Even at the 
matrix model level, the definition of the quantum hamiltonian has ordering ambiguities. The first two
traces, in terms of the matrix elements of $P$,
\be
\tr P = \sum_n P_{nn} ~,~~~ \tr P^2 = \sum_{m,n} P_{mn} P_{nm}
\ee
are uniquely defined. At the cubic level, however, we already see that there are two possible orderings:
\be
\tr P^3 = \sum_{m,n,l} P_{mn} P_{nl} P_{lm} ~~~ {\rm or} ~~~
\sum_{m,n,l} P_{mn} P_{lm} P_{nl}
\ee
The two are classically the same but quantum mechanically inequivalent, differing by lower-order
terms. Although the first leads to the conventionally defined Casimir, there is no reason not to
consider the second. In fact, the sum of the two leads to an expression where the first subleading
correction to $\sum p_k^3$ cancels.

We see that the exact definition of the hamiltonian depends on the ordering of its terms.
This is not surprising, since the original field action contained infinitely high powers of time derivatives
and such terms require a precise ordering.

We can see this ambiguity at the field theory level before we reduce
to the matrix model by a method analogous to the heat kernel in standard two-dimensional Yang Mills:
we tesselate spacetime into small plaquettes of arbitrary shape and size and perform the euclidean
path integral over the gauge fields inside each plaquette. From gauge invariance, the result for each
plaquette will only depend on the
holonomy (Wilson loop) $W$ of the gauge field around the plaquette.
The fixed-point expression must be of the form
\be
Z = \sum_R d_R \, e^{-A E_R } \chi_{_R} (W)
\label{pl}
\ee
with $A$ the area of the plaquette, $d_R$ the dimension of irrep $R$ and $E_R$ a number depending only on $R$.
We can then consider two adjacent plaquettes with Wilson element $U$ on their common boundary and holonomies
$W_1 U$ and $U^{-1} W_2$ ($W_1$ and $W_2$ being the non-common parts) and integrate
their path integrals $Z_1 (UW_1 ) Z_2 (U^{-1} W_2)$ over the common part $U$ 
to calculate the path integral for the combined plaquette.
Due to the orthogonality property of the irreps
\be
\int [dU] \, \chi_R (WU) \chi_{R'} (U^{-1} V) = \frac{1}{d_R} 
\, \delta_{R R'} \, \chi_R (WV)
\ee
we see that the result will be of the form
\be
Z_{12} = \sum_R d_R \, e^{-(A_1 + A_2) E_R } \chi_{_R} (W_1 W_2 )
\ee
involving their total area and the holonomy $W_1 W_2$ around the total plaquette, verifying the
consistency of the expression (\ref{pl}) for the fixed-point
partition function.

The only extra requirement is that the above expression be the quantization of a specific classical
action. For this, we need to ensure that for $A \to 0$ the expression in the exponent of (\ref{pl})
goes over to the classical action for the fields. For small $A$, the sum in (\ref{pl}) is dominated by
large irreps, that is, by large values of $p_k$ in the fermionic description. For such irreps, we must
have
\be
E_R = \sum_{k=1}^N\sqrt{\lambda ^2 + T^2 p_k^2} ~~~{\rm for} ~~ |p_k | \gg N
\label{ER}\ee
{\it Any} $E_R$ with the above property provides a consistent gauge invariant quantization of
the same Born-Infeld classical field teory. Choosing the expression (\ref{expand}) in terms of
the standard Casimirs in just one of many possibilities.

We can, therefore, adopt the simplest definition in which the expression (\ref{ER}) holds for {\it all}
irreps and define the hamiltonian as
\be
H = \sum_{n=1}^N E( p_n ) =L \sum_{n=1}^N\sqrt{\lambda ^2 + T^2 p_n^2}
\ee 
The above assigns a positive value to the energy of the ground state, which is inconsequential
for expectation value calculations and can easily be removed.
In some sense, the above is the most natural definition, since the dynamics of the gauge field reduce
to those of a set of uncoupled relativistic particles, admitting the interpretation of points on
a relativistic brane in a dual description.

Our final result is that Born-Infeld gauge theory on the cylinder reduces to a set of relativistic fermions
on a dual circle. If we incorporate a factor of $LT$ in the momentum, the energy expression
for each particle becomes
\be
{\tilde E} ({\tilde p} ) =  \sqrt{\lambda ^2 L^2 + {\tilde p}_n^2} ~,~~~ {\tilde p} = LT p
\ee
representing a particle of mass $\lambda L$ on a circle of radius $R = (TL)^{-1}$.
This is to be contrasted to regular two-dimensional Yang Mills on the cylinder, which is equivalent
to a set of nonrelativistic fermions and in which there is no unique identification of particle mass
and radius of the dual circle, the two appearing as one overall coefficient.

\section{The large-$N$ limit and string description}

There are various ways to take the large-$N$ limit in the above theory. The one relevant to the
string interpretation is what we can call the conformal field theory limit. In this limit, the
low-lying energy excitations of the theory become equally spaced and approach those of a
$c=1$ conformal field theory, that is, a relativistic fermion.

The excitations of the BI theory consist of fermion excitations above the Fermi level. There are
two Fermi levels, at $p = \pm (N-1)/2$, leading to two left- and a right-moving non-interacting sectors.
(Depletion of the Fermi sea corresonds to nonperturbative in $1/N$ interaction effects.)
Concentrating on excitations near the right-moving Fermi level $p_{_F} = (N-1)/2$, 
a fermion excited from
$p= (N-1)/2 - m$ to $(N-1)/2 + n$, with $m,n$ positive and of order 1, has excitation energy
\be
\Delta E = E(p_{_F} - m ) - E(p_{_F} +n ) \simeq \partial_p (p_{_F} ) (n+m) 
~~{\rm (for~}N\gg1{\rm )}
\ee
So the scale of the energy gap is set by the Fermi velocity (velocity of sound on dual space)
\be
v_{_F} = f(p_{_F}) ~, ~~~ f(p) \equiv \frac{\partial E(p)}{\partial p}
= \frac{L T^2 p}{\sqrt{\lambda ^2 + T^2 p^2}}
\ee
For $p_F = (N-1)/2$ the above will be of order $N^0$ if $T$ does not
scale but $\lambda$ scales with $N$, that is,
\be
\lambda = \lambda_o N
\label{lN}
\ee
From (\ref{YMg}) we see that in the Yang-Mills limit the above scaling is cosistent
with the standard 't Hooft scaling, with 't Hooft coupling
\be
g_o^2 = \frac{T^2}{\lambda_o}
\ee
From now on we will always assume the expression (\ref{lN}) for $\lambda$ and will write
$\lambda$ instead of $\lambda_o$ to alleviate notation.

The string picture of the gauge theory on the cylinder remains largely as in standard Yang-Mills:
the leading-$N$ terms in the excitation energy represent a theory of free strings
wrapping around $S^1$ with string tension
\be
T_{st} = \frac{v_{_F}}{L} = \frac{T^2}{\sqrt{4\lambda^2 + T^2}}
\ee
The term of order $n^2$ in the expansion of $E( p_F +n) - E( p_F )$ is a $1/N$ correction that
introduces a cubic string interaction representing string splitting or joining. The string coupling
constant is
\be
g_{st} = \frac{1}{2} \frac{\partial^2 E( p_{_{F}} )}{\partial p^2} =
\frac{\lambda^2 T^2}{N \left(4\lambda^2 + T^2\right)^{\frac{3}{2}}}
\ee
The difference from Yang-Mills theory comes from the existence of higher orders in the expansion
of $E(p)$ in $p$, which are absent in the Yang-Mills case. The cubic term, of order $N^{-2}$,
introduces a quartic string interaction that represents a localized double string interaction, that is,
two pairs of strings touching and reconnecting at the same point of space and time. Such interactions
are not so natural from the worldsheet point of view. Higher terms lead to higher yet order string
interactions. Overall, we have a nonpolynomial string field theory.

\section{Large $N$ phase transition on the sphere}

The partition function on the circle is given by the path integral on a euclidean torus $(L,\beta )$.
The result is
\be
Z = \sum_{\{ p_k \}} e^{-A \sum_n \sqrt{\lambda^2 N^2 + T^2 p_n^2 }}
\ee
with $A = L\beta$ the area of the wordsheet. The summation is over all combinations of
fermionic momenta $p_1 > p_2 > \cdots p_N$ on a (half) integer lattice for (even) odd $N$.

For spacetimes of genus $g$, the partition function is similar but with an extra measure factor $d_R^{2-2g}$ in each term. This factor can be understood as a remnant from the plaquette formula (\ref{pl}) as we
coalesce the plaquettes on spacetimes of different topologies.
For spherical topology, in particular, it can be understood as
arising from the insertion of a singular wavefunction at the north and
south pole of the sphere, representing the constraint $W=1$ at these
points in a canonical formulation \cite{MiPo2}. 
In terms of fermion momenta the extra factor $d_R^2$ on the sphere
is expressed as a Vendermonde-like product
\be
d_R^2 =  \frac{\prod_{n<m} (p_n - p_m )^2}{\prod_{n<m} (n - m)^2}
\ee
The denominator is the product for the ground state momenta, ensuring
$d_R = 1$ for the singlet, and contributes an overall normalization
factor that will be omitted.
The resulting partition function on the sphere is
\be
Z_{sph} = \sum_{\{ p_k \}} \prod_{n<m} (p_n - p_m )^2 e^{-A \sum_n \sqrt{\lambda^2 N^2 + T^2 p_n^2 }}
= \sum_{\{ p_k \}} e^{-S_{eff}}
\ee
The effective action contains the exponentiated measure and reads
\be
S_{eff} = A \sum_n \sqrt{\lambda^2 N^2 + T^2 p_n^2 } - 2 \sum_{n<m}
\ln | p_n - p_m |
\ee
The measure introduces a repulsive logarithmic two-body potential in the momenta.

For large $N$ the effective action is of order $N^2$, since $p_n$ are of order $N$, and the partition will be dominated by the classical
minimal effective energy configuration
in a saddle-point approximation. By differentiating $S_{eff}$ with respect to $p_n$ we obtain the
minimal energy condition
\be
A \frac{T^2 p_n}{\sqrt{\lambda^2 N^2 + T^2 p_n^2}} - \sum_{m(\neq n)} \frac{2}{p_n - p_m } =0
\ee
For large $N$ we can approximate the distribution of momenta with a continuous density ${\tilde \rho} (p)$.
The minimum energy condition becomes
\be
\frac{A}{2} \frac{T^2 p}{\sqrt{\lambda^2 N^2 + T^2 p}}
= \dashint \frac{{\tilde \rho}( p' )}{p- p'} dp'  ~~~ {\rm for}~~ 
{\tilde \rho}(p) \neq 0
\label{barro}
\ee
with $\tilde \rho$ satisfying
\be
{\tilde \rho} \geq 0 ~,~~~ \int {\tilde \rho} (p) dp = N
\ee
We can define a rescaled variable $x = p/N$ and a corresponding density
\be
\rho (x) = {\tilde \rho} ( Nx ) ~,~~~ \int \rho(x) dx = 1
\ee
In terms of the new variable and density the condition (\ref{barro}) becomes
\be
\frac{A}{2} \frac{T^2 x}{\sqrt{\lambda^2 + T^2 x^2}}
= \dashint \frac{\rho ( y )}{x-y} dy  ~~~ {\rm for}~~ \rho (x) \neq 0
\label{eqrho}
\ee
eliminating all reference to $N$ in the large-$N$ limit.

The solution of the above equation is well-known. The function
\be
u(z) = \int \frac{\rho(y)}{z-y} dy
\ee
is analytic on the upper half plane and behaves as $\rho (z) \sim 1/z$ at $z \to \infty$. Near the real
axis it becomes
\be
u(x+i0) = -i\pi \rho (x) + \dashint \frac{\rho ( y )}{x-y} dy 
\ee
For a symmetric distribution $\rho(-x) = \rho(x)$ that vanishes outside of an interval $(-a,a)$
the solution for $u(z)$ is
\be
u(z) = \frac{1}{2\pi i} \sqrt{a^2 - z^2} \oint \frac{A T^2 s}{2(s-z)\sqrt{\lambda^2 + T^2 s^2} \sqrt{a^2 - s^2}} ds
\label{u}
\ee
In the above the square roots are defined with a cut along $(-a,a)$ and the integration controur is
clockwise around the cut but not including the pole at $s=z$. It is easy to see from the above formula that $u(x+i0)$
is real for $|x|>a$, while its imaginary part is the left hand side of (\ref{eqrho}) for $|x| <a$. Therefore it
satisfies (\ref{eqrho}), provided it also has the proper asymptotic behavior for large $z$. This will be
ensured if
\be
\int_{-a}^a \frac{A T^2 s^2}{\sqrt{\lambda^2 + T^2 s^2} \sqrt{a^2 - s^2}} ds = 2\pi
\label{z1}\ee
The density $\rho(x)$ is then recovered as
\be
\rho(x) = \frac{1}{\pi^2} \sqrt{a^2 - x^2} ~ \dashint_{-a}^a 
\frac{A T^2 s}{2(s-x)\sqrt{\lambda^2 + T^2 s^2} \sqrt{a^2 - s^2}} ds
\ee
For the case of standard Yang-Mills, where $f(p) = A g^2 p^2 /2$,
the corresponding integral (\ref{u}) can be easily calculated by
blowing up the contour to infinity, and leads to the Wigner semicircle
distribution. In our case this is not so easy, since the contour
encounters the cut of the square root in the denominator on the
imaginary axis $(i \lambda/T , \infty )$. The integrals can be
expressed in terms of elliptic functions and implicitly define $a$ and $\rho(x)$.

The above solution is valid as long as $A$ is not too big, in the so-called weak coupling phase. As $A$ increases,
$a$ decreases and the momentum distribution becomes denser. The momenta, however, are fermionic and lie
on a lattice of spacing 1, so their density cannot exceed 1. Correspondingly, $\rho(x) = {\tilde \rho} (Nx)$
cannot exceed 1. The maximum of $\rho(x)$ occurs at $x=0$. Therefore, when $\rho(0)$ reaches the value 1,
fermionic momenta will start condensing and the above solution will not be valid any more, signaling a phase
transition. To find the critical area we put $\rho(0) = 1$:
\be
\rho(0) = \frac{a A_{cr} T^2}{2\pi^2}  \int_{-a}^a 
\frac{ds}{\sqrt{\lambda^2 + T^2 s^2} \sqrt{a^2 - s^2}} = 1
\label{rho1}
\ee
Combining (\ref{z1}) and (\ref{rho1}) we obtain an equation for the width $a$ at critical area:
\be
\int_{-a}^a  \frac{a - \pi s^2}{\sqrt{\lambda^2 + T^2 s^2} \sqrt{a^2 - s^2}} ds = 0
\ee
which fixes $a$ in terms of $T/\lambda$ and, upon inserting in 
(\ref{rho1}), it determines $A_{cr}$.

For $A > A_{cr}$ the solution develops a flat central part where the
fermion momenta condense, and an outer tail part:
\begin{eqnarray}
\rho (x) &=  1 & |x| < b \cr
&~~~=  {\bar \rho} (x) &  b < |x| < a \cr
&=  0 & |x| >a
\end{eqnarray}
The contribution from the flat central part can be taken explicitly
into account in the equation (\ref{eqrho}) producing an extra
logarithmic potential \cite{DoKa}. The remaining density
${\bar \rho} (x)$ vanishing outside $(-a,a)$ and inside $(-b,b)$
can be found in a way similar to $\rho (x)$.
The solution for its analytic extension ${\bar u} (z)$ becomes a
two-cut integral with an additional logarithm cut between $(-b,b)$:
\be
{\bar u} (z) = \frac{1}{2\pi i} \sqrt{(a^2 - z^2 )(b^2 - z^2)} 
\oint \frac{ \frac{A T^2 s}{2\sqrt{\lambda^2 + T^2 s^2}}
+ \ln\frac{s-b}{s+b}}
{(s-z)\sqrt{(a^2 - s^2 )(b^2 - s^2 )}} ds
\label{uu}
\ee
with the contour encircling the square root cuts between
$(-a,-b)$ and $(b,a)$ but not the log cut and the pole at $z$.
The second part of the above integral, involving the logarithm,
can be explicitly
evaluated by deforming the contour around the log cut (we
encounter the pole at $s=z$ and no other cuts). The result is
\beqs
&&\frac{1}{2\pi i} \sqrt{(a^2 - z^2 )(b^2 - z^2)} 
\oint \frac{ \ln\frac{s-b}{s+b}}
{(s-z)\sqrt{(a^2 - s^2 )(b^2 - s^2 )}} ds = \cr
&&\ln \frac{z-b}{z+b} - \sqrt{(a^2 - z^2 )(b^2 - z^2)}
\int_{-b}^b \frac{ds}
{(s-z)\sqrt{(a^2 - s^2 )(b^2 - s^2 )}} ds
\eeqs
For $z=x+i0$ the imaginary part of the logarithm above vanishes
for $|x| >b$ and equals $i\pi$ for $|x| <b$. It thus contributes
$-1$ to the density ${\bar \rho} (x)$ in the interval $(-b,b)$
and zero outside. Therefore, removing it restores the density
to its full value $\rho (x)$ (equal to 1 between $-b$ and $b$).
The density is reproduced by the above integral plus the first
part (non-logarithm) of the integral in (\ref{uu}).
Taking also into account the even nature of $\rho (x)$ we obtain
\be
\rho (x) = \frac{|x|}{\pi^2} \sqrt{(a^2 - x^2 )(x^2 - b^2)}
\left[
\dashint_b^a \frac{A T^2 s}{\sqrt{\lambda^2 + T^2 s^2}} +
\int_{-b}^b \pi \right]
\frac{ds}{(s^2-x^2)\sqrt{(a^2 - s^2 )|b^2 - s^2 |}}
\ee
(Note that in the case of standard Yang-Mills the first integral above
vanishes, as can be shown by contour integration, and only the second
term, arising from the logarithm integral, survives.)
We must also ensure the proper asymptotic behavior of ${\bar u}(z)$
at infinity, that is,
\be
u(z) = 0 \cdot z + 0 \cdot 1 + \frac{1-2b}{z} + O(z^{-2} )
\ee
The vanishing of the constant term above is an identity, but
the terms of order $z$ and $z^{-1}$ give two conditions that
fix, in principle, $a$ and $b$ in terms of $A$.

\section{The limit $\lambda /T \to 0$}

The case $\lambda \ll T$ is of particular interest: 
in terms of fermions,
the dispersion relation becomes linear and the fermions
become massless. The Fermi velocity is constant and the
corresponding string theory contains no higher order terms and
becomes free. In a sense, this is the ``stringiest" version of
gauge theory and does not even require a large-$N$ limit to
manifest a perturbative free string behavior. Finite-$N$
effects arise only as nonperturbative corrections.

It is interesting that in this case the eigenvalue distribution
can be calculated exactly. Putting $\lambda = 0$ and substituting
$\sqrt{\lambda^2 + T^2 x^2}$ by $T|x|$ we obtain integrals with a
cut along the entire imaginary axis that can be explicitly evaluated.
The normalization condition (\ref{z1}) gives
\be
a = \frac{\pi}{AT}
\ee
while the expression for $\rho$ gives
\be
\rho(x) = \frac{AT}{\pi^2} \ln \frac{\pi + \sqrt{\pi^2 - A^2 T^2 x^2}}
{AT |x|}
\label{weak}
\ee
We see that $\rho(0) = \infty$ for all $A$, and so the model is always in the
strong coupling (stringy) phase. The above solution, therefore, is not
really valid but we must instead calculate the two-cut solution with a
flat central region for $\rho(x)$. It is still a good approximation to the exact solution
for small enough $A$, that is, $AT \ll 1$. In that case the solution for the density
is the above, for $|x|>b$, and 1 for$|x| <b$, with $b$ the value for which the above
function reaches the value 1, that is,
\be
\frac{1}{b} = \frac{AT}{\pi} {\rm ch} \frac{\pi^2}{AT}
\label{b}
\ee
The above is clearly nonperturbative in $A$. For small areas, the solution for $\rho(x)$
differs very little from the would-be weak coupling solution (\ref{weak}). For $AT \gg 1$,
on the other hand, the solution approaches a true ``stringy" state of a fully filled Fermi sea
with few momenta spreading above the Fermi levels and becomes identical to the corresponding
Yang-Mills solution upon identifying the Fermi velocities, or
\be
2T = g^2
\ee
with $g$ the 't Hooft Yang-Mills coupling ($g_o$ in (\ref{YMN})).

For small nonzero values of $\lambda$ ($\lambda \ll T$) we can estimate
the critical area $A_{cr}$. The integral in (\ref{z1})
is of order $\lambda^0$ while the integral (\ref{rho1}) has a
logarithmic divergence in $\lambda$. To leading order we obtain
\be
\pi a = \ln \frac{2Ta}{\lambda}
\ee
which has as leading log solution
\be
a = \frac{1}{\pi} \ln \frac{2T}{\pi \lambda}
\ee
Altogether this gives the critical area
\be
A_{cr} = \frac{\pi^2}{T \ln \frac{2T}{\pi \lambda}}
\ee
which is, again, nonperturbative in $\lambda$. The same conclusion
can be reached by putting $b = \lambda/T$ in formula (\ref{b}),
which is the value of $s$ for which the two terms in the expression
$\sqrt{\lambda^2 + T^2 s^2}$ become comparable and thus $\lambda$
starts regulating the behavior at $s=0$. For $b$ less than that we
do not expect $\rho (0)$ to reach 1, so at this value of $b$ in
(\ref{b}) we expect a phase transition.

Finally, we can calculate the free energy ${\cal F}(A)$ in the weakly coupled case of small area.
It will be given by the value of the effective action for the saddle point distribution
for $p_n$. Since $\partial S_{eff} /\partial p_n = 0$ at the classical saddle point, 
we have
\be
\frac{\partial {\cal F}}{\partial A}  = \frac{\partial}{\partial A} S_{eff} 
= \sum_n T | p_n | = T \int |p| \, {\tilde \rho} (p) \, dp
= N^2 T \int |x| \, \rho(x) \, dx
\ee
The density $\rho(x)$ is given by (\ref{weak}) up to nonperturbative corrections in $A$.
An explicit calculation gives
\be
\frac{\partial {\cal F}}{\partial A}  = \frac{1}{A} ~~ \rightarrow~~ {\cal F} = \ln A
\ee
up to a constant. We should also subtract the ground state energy of the fermions,
such that the vacuum have zero energy. For the ground state momenta (\ref{pgr})
and the large $N$ limit we have
\be
\beta E_o = AT \sum_n | p_{n,o} | = \frac{1}{4} AT N^2
\ee
so overall the free energy is
\be
{\cal F} = N^2 \left( \ln A - \frac{1}{4} TA \right)
\ee
up to an $A$-independent constant. The $\ln A$ part is essentially fixed by the scaling
properties of the fermion particle energy. An expression $E(p) \sim p^\alpha$ would contribute
a term $\alpha^{-1} \ln A$. In the case of Yang-Mills we have $\alpha =2$, while in
our case $\alpha = 1$.

\section{Conclusions}

The properties of Born-Infeld two-dimensional gauge theory in general parallel those of standard
Yang-Mills, with some interesting twists. The disparity between the two becomes apparent for large
values of the gauge field, as expected. On the cylinder, large energy excitations tend to preserve
their linear dispersion relation over a wider range, although the deviations are non-polynomial.
On the sphere, a phase transition also occurs, but the critical area decreases as the Born-Infeld
theory becomes more relativistic and vanishes in the tensionless limit $\lambda \to 0$. In that
limit, the theory on the cylinder becomes a free string theory, receiving only nonperturbative
corrections in the large-$N$ limit.

There are many issues that remain to be investigated. The expansion of the free energy as a function
of the area and its nonanalyticity near the transition point on the sphere could be examined with a
view to clarify the stringy nature of the strong coupling phase. The question of $U(1)$ sectors
is also an open one: a global momentum shift of the fermions is in principle a low energy excitation but,
in the large-$N$ limit, it becomes nonperturbative. The evaluation of the partition function can be
performed around an isolated $U(1)$ (total momentum) sector, similar to the Yang-Mills case \cite{MiPo2}.

Finally, the calculation and behavior of Wilson
loop expectation values is a very interesting issue. In the case of the sphere they would probe the
nature of the phase transition, and of the validity of the string description. On the cylinder, insertion of (one or several) timelike
Wilson loops would promote the fermion system into an interacting one with internal degrees of
freedom, which would constitute integrable and solvable many-body systems \cite{BlLa,Pol2}.
The obvious conjecture would be that these systems are genaralizations of the Ruijsenaars-Schneider
system of `relativistic' fermions including internal degrees of freedom, but the exact form of
the hamiltonian has to be worked out.

\eject


\begin{thebibliography}{99}

\bibitem{Hoo} 
  G.~'t Hooft,
  Nucl.\ Phys.\ B {\bf 72}, 461 (1974).

\bibitem{Wil} 
  K.~G.~Wilson,
  Phys.\ Rev.\ D {\bf 10}, 2445 (1974).

\bibitem{Sak} B.~Sakita,
  Phys.\ Rev.\ D {\bf 21}, 1067 (1980).

\bibitem{BBHP} 
  W.~A.~Bardeen, I.~Bars, A.~J.~Hanson and R.~D.~Peccei,
  Phys.\ Rev.\ D {\bf 13}, 2364 (1976).

\bibitem{GrWi} 
  D.~J.~Gross and E.~Witten,
  Phys.\ Rev.\ D {\bf 21}, 446 (1980).

\bibitem{Wad} 
  S.~R.~Wadia,
  Phys.\ Lett.\ B {\bf 93}, 403 (1980).

\bibitem{Mig} 
  A.~A.~Migdal,
  Sov.\ Phys.\ JETP {\bf 42}, 413 (1975)
  [Zh.\ Eksp.\ Teor.\ Fiz.\  {\bf 69}, 810 (1975)].

\bibitem{Rus} 
  B.~E.~Rusakov,
  Mod.\ Phys.\ Lett.\ A {\bf 5}, 693 (1990).

\bibitem{KaKo} 
  V.~A.~Kazakov and I.~K.~Kostov,
  Nucl.\ Phys.\ B {\bf 176}, 199 (1980);
  Phys.\ Lett.\ B {\bf 105}, 453 (1981).

\bibitem{Kaz} 
  V.~A.~Kazakov,
  Nucl.\ Phys.\ B {\bf 179}, 283 (1981).

\bibitem{Kos} 
  I.~K.~Kostov,
  Phys.\ Lett.\ B {\bf 138}, 191 (1984).

\bibitem{Gro} 
  D.~J.~Gross,
  Nucl.\ Phys.\ B {\bf 400}, 161 (1993)
  [hep-th/9212149].

\bibitem{Min} 
  J.~A.~Minahan,
  Phys.\ Rev.\ D {\bf 47}, 3430 (1993)
  [hep-th/9301003].

\bibitem{GrTa} 
  D.~J.~Gross and W.~Taylor,
  Nucl.\ Phys.\ B {\bf 400}, 181 (1993)
  [hep-th/9301068];

\bibitem{DoKa} 
  M.~R.~Douglas and V.~A.~Kazakov,
  Phys.\ Lett.\ B {\bf 319}, 219 (1993)
  [hep-th/9305047].

\bibitem{MiPo1} 
  J.~A.~Minahan and A.~P.~Polychronakos,
  Phys.\ Lett.\ B {\bf 312}, 155 (1993)
  [hep-th/9303153].

\bibitem{MiPo2} 
  J.~A.~Minahan and A.~P.~Polychronakos,
  Nucl.\ Phys.\ B {\bf 422}, 172 (1994)
  [hep-th/9309119].

\bibitem{Pol1} 
  A.~P.~Polychronakos,
  Phys.\ Lett.\ B {\bf 266}, 29 (1991).

\bibitem{DaJe} 
  S.~R.~Das and A.~Jevicki,
  Mod.\ Phys.\ Lett.\ A {\bf 5}, 1639 (1990).

\bibitem{GoNe} 
  A.~Gorsky and N.~Nekrasov,
  Nucl.\ Phys.\ B {\bf 414}, 213 (1994)
  [hep-th/9304047].

\bibitem{MiPo3} 
  J.~A.~Minahan and A.~P.~Polychronakos,
  Phys.\ Lett.\ B {\bf 326}, 288 (1994)
  [hep-th/9309044].

\bibitem{FrTs} 
  E.~S.~Fradkin and A.~A.~Tseytlin,
  Phys.\ Lett.\ B {\bf 163}, 123 (1985).

\bibitem{Lei} 
  R.~G.~Leigh,
  Mod.\ Phys.\ Lett.\ A {\bf 4}, 2767 (1989).

\bibitem{Tse} 
  A.~A.~Tseytlin,
  Nucl.\ Phys.\ B {\bf 501}, 41 (1997)
  [hep-th/9701125].

\bibitem{Dou} 
  M.~R.~Douglas,
  hep-th/9303159 and hep-th/9311130.

\bibitem{BlLa} 
  J.~Blom and E.~Langmann,
  Phys.\ Lett.\ B {\bf 429}, 336 (1998)
  [solv-int/9804007].

\bibitem{Pol2} 
  A.~P.~Polychronakos,
  Nucl.\ Phys.\ B {\bf 546}, 495 (1999)
  [hep-th/9806189];
  Nucl.\ Phys.\ B {\bf 543}, 485 (1999)
  [hep-th/9810211].


\end{thebibliography}
\end{document}